\documentclass[sigconf]{acmart}

\renewcommand\footnotetextcopyrightpermission[1]{} 
\AtBeginDocument{%
  \providecommand\BibTeX{{%
    \normalfont B\kern-0.5em{\scshape i\kern-0.25em b}\kern-0.8em\TeX}}}

\begin{document}

\title{Maxwell: a hardware and software highly integrated compute - storage system}

\author{Ling.~Ma}
\authornotemark[1]
\affiliation{%
  \institution{Ant Group}
  \country{China}
}
\email{ling.ml@antgroup.com}

\author{Wei.~Zhou}
\affiliation{%
  \institution{Ant Group}
  \country{China}
}
\email{eric.zw@antgroup.com}

\author{Sihai.~Yao}
\affiliation{%
  \institution{Ant Group}
  \country{China}
}
\email{sihai.ysh@antgroup.com}

\author{Hong.~Huang}
\affiliation{%
  \institution{Ant Group}
  \country{China}
}
\email{hh.hh@antgroup.com}

\author{Yinsong.~Xie}
\affiliation{%
  \institution{Ant Group}
  \country{China}
}
\email{yinsong.xys@antgroup.com}

\author{Haohao.~Liu}
\affiliation{%
  \institution{Ant Group}
  \country{China}
}
\email{yaohao.lhh@antgroup.com}

\author{Changhua.~He}
\affiliation{%
  \institution{Ant Group}
  \country{China}
}
\email{changhua.hch@antgroup.com}

\renewcommand{\shortauthors}{Ling,Ma and Wei,Zhou, et al.}

\begin{abstract}
The compute-storage framework is responsible for data storage and processing, and acts as the digital chassis of all upper-level businesses. The performance of the framework affects the business's processing throughput, latency, jitter, and etc., and also determines the theoretical performance upper bound that the business can achieve. In financial applications, the compute-storage framework must have high reliability and high throughput, but with low latency as well as low jitter characteristics. For some scenarios such as hot-spot account update, the performance of the compute-storage framework even surfaces to become a server performance bottleneck of the whole business system.
\newline In this paper, we study the hot-spot account issue faced by Alipay and present our exciting solution to this problem by developing a new compute-storage system, called Maxwell. Maxwell is a distributed compute-storage system with integrated hardware and software optimizations. Maxwell does not rely on any specific hardware (e.g. GPUs or FPGAs). Instead, it takes deep advantage of computer components' characteristics, such as disk, network, operating system and CPU, and aims to emit the ultimate performance of both hardware and software. In comparison with the existing hot-spot account updating solutions deployed online, Maxwell achieves three orders of magnitude performance improvement for end-to-end evaluation. Meanwhile, Maxwell also demonstrates remarkable performance gains in other related businesses of Ant Group.
\end{abstract}


\setcopyright{none}

\maketitle

\pagestyle{plain}

\section{Introduction}
With the increasing number of users and the increasing diversity of business scenarios, the influence of poor updating performance on hot-spot account is gradually highlighted. A hot-spot account is an account that generates a large number of transaction requests in a short period of time. The duration may be only a few seconds, or it may last for a longer time. This hot-spot account problem is mostly encountered on the accounts of big sellers such as livestreamers and shops that participate in shopping festivals. In this scenario, the number of read and write requests to a single account can be very big, generating a continuous flow peak. If the accounting system is incapable of processing these requests, it not only affects the user's payment experience, but also may result in a wide range of system failures, or even trigger a cluster avalanche.

To address the challenge of updating on hot-spots, the industry generally uses methods like account splitting, buffering, summaring, transaction limiting and even configuring hot-spot accounts as non-real-time ones. However, these methods bring new problems as side-effects, stated as follows.
\begin{itemize}
\item Non-real-time accounting may result in overdrafts and capital losses.
\item Account balance updates are not real-time, causing poor collection experience.
\item High labor costs, as a lot of preparation work is demanded to be done manually before any account is configured as a hot-spot account.
\item Once an exception occurs, recovery costs are high and can lead to customer churn.
\end{itemize}

Having investigated existing solutions, we have decided to develop a new computing storage system that is compliant with the performance indicators of the hot-spot account updating business. This system originally targets a dedicated hot-spot accounting database and has the following characteristics:
\begin{itemize}
\item \textbf{High reliability.} Obviously, any financial database should be guaranteed to be highly reliable.
\item \textbf{High throughput.} Facing and solving the flow peak directly, the scheme should be simple and have no side effects.
\item \textbf{Low latency.} Account updating is only one process in the synchronization path of the long and complex payment process. There are strict requirements and control over processing delay, which can cause payment failure and a series of consequences.
\item \textbf{Low jitter.} Not only on processing throughput but also on processing delay. High jitter will bring shock to upstream and downstream processing, resulting in request timeouts, accumulating, and other issues.
\item \textbf{No inflection point.} Even if the amount of simultaneous requests exceeds the maximum throughput of the compute-storage system, there must be no performance inflection point that can result in a rapid deterioration of the entire link performance or even an avalanche.
\item \textbf{Support transactions.} Ensure the atomicity of capital inflow and outflow operations.
\item \textbf{Linear expansion.} When the available resources in the system increase, the system throughput can increase linearly.
\end{itemize}

Furthermore, we abstract and generalize the hot-spot accounting database to derive Maxwell distributed compute-storage system, which has better versatility. We will discuss the design and implementation of Maxwell distributed compute-storage system in detail.

\section{Maxwell Architecture}
Due to its good performance, Maxwell hot-spot accounting database has increased its account number running on from thousands to hundreds of millions, and expanded its application scenarios to cover several aspects of financial services. Maxwell not only ensures a good user experience, but also saves business costs. Through continuous updates, it has evolved from a dedicated system for hot-spot accounts to a common distributed compute-storage system that can quickly build solutions that serve other business areas while inheriting good performance characteristics. The following sections describe the architecture of Maxwell compute-storage.

\begin{figure*}
	\centering
	\includegraphics[width=\linewidth]{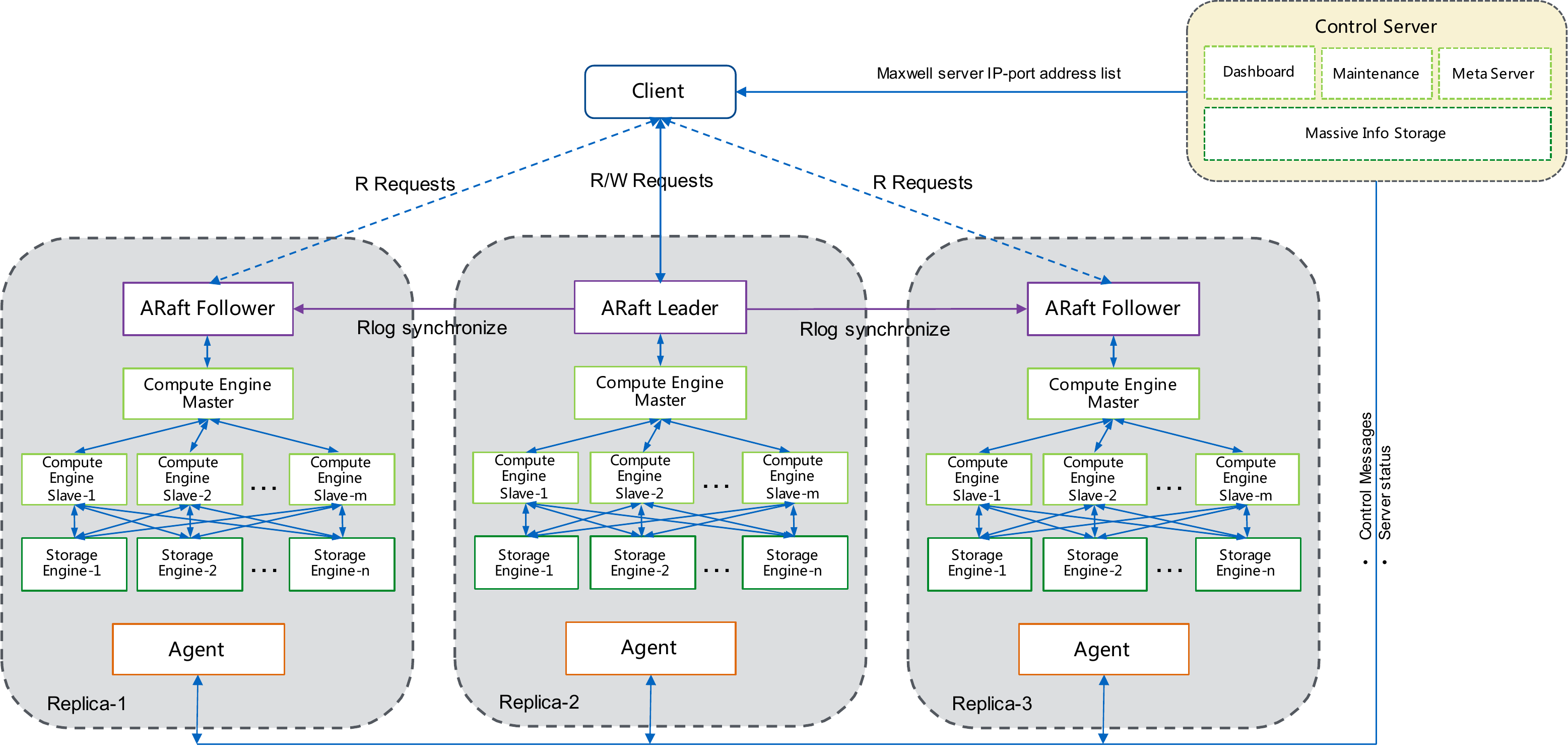}
	\caption{Architectural overview of Maxwell.}
	\label{fig:fig1}
\end{figure*}

Figure \ref{fig:fig1} illustrates the Maxwell's compute storage architecture, which consists of seven components: client, ARaft, computing engine, storage engine, agent, metaserver and web console.
\begin{itemize}
\item \textbf{Client.} Interacts with Maxwell servers, and provides for user integration.
\item \textbf{ARaft.} Maintains data consistency between multiple replicas. When an RLog is received by most replicas, it is sent to the compute engine for execution.
\item \textbf{Computing engine.} Processes user requests, and computes data processing, load balancing, etc. The number of compute engines can be expanded according to business's demands.
\item \textbf{Storage engine.} Provides basic data storage and query service. The number of storage engines can be scaled to fit business demands.
\item \textbf{Agent. }
	\begin{itemize}
	\item Interacts with MetaServer to collect replica status information.
	\item Performs predefined commands, such as starting a service process.
	\end{itemize}
\item \textbf{MetaServer.}
	\begin{itemize}
	\item Offer maintenance and management on service cluster
	\item Provides service discovery to clients
	\end{itemize}
\item \textbf{Web console.} Provides an operating interface for operations and maintenance(O \& M) administrators.
\end{itemize}

As shown in Figure \ref{fig:fig1}, a Maxwell cluster contains multiple replicas at least three to make logic and function complete. Each replica mainly consists of three core components: storage engine, compute engine, and ARaft. The MetaServer maintains information about all clusters and their replicas, including the name of the cluster, the number of replicas, the IP addresses and ports of each replica. The client obtains the IP address and port of each replica of the cluster to be accessed through the MetaSever, and then connects to it. All write requests from clients are packaged into RLogs for storage, distribution, and execution in the order that ARaft Leader receives. 

ARaft in multiple replicas uses a set of protocols to ensure multiple compute engine executes the same client requests in the same order, thus ensuring data consistency and high availability among multiple replicas. Typically, three or five centers are used per cluster for deployment. The number of compute engines and storage engines in replicas can be expanded according to business requirements, and they can run on the same server or on multiple servers. The components are described separately below.

\section{Storage Engine}
Maxwell storage engine provides ordinary KV and timed KV data storage and query services. It mainly provides the following functions.
\begin{itemize}
\item \textbf{Data addition, deletion, modification, and checking based on time / sequence dimensions.} The data is stored in the database in chronological order and related attributes (TAGs).
\item \textbf{A variety of query methods.}
	\begin{itemize}
	\item Time/Attribute KV mode: retrieve data in forward/backward order.
	\item Interval mode: use time/attribute interval and table information to query single-column and multi-column data.
	\end{itemize}
\item \textbf{Data insertion}
	\begin{itemize}
	\item Out of order insertion.
	\item Single / batch insertion.
	\item Data updating.
	\end{itemize}
\item \textbf{Hot data cache.} Hot data is buffered in the memory, and build-in strategies define how to evict and flush data.
\item \textbf{Two-phased transactional operation support.} Modifications to data are only visible after commit or discarded when rollback. 
\end{itemize}

\subsection{Data Model}
A Maxwell storage engine can manage several databases, where each database contains several tables, and each table has several rows of data, and each row owns several columns. The number of columns contained in a table is specified at the time when the table is created. Column dynamic expansion is not supported currently. Supported column data types include Integer, Long, Double, Binary (variable-length binary data), etc. For a table with multiple columns, it's required to provide all the column data when inserting a row. The data is arranged in ascending order of the Key in the table, and the indexing information of the Key is also maintained in the same table.
\begin{itemize}
\item \textbf{Data insertion.} The storage engine first searches the table to see whether the key exists or not.  An insertion operation is conducted if it does not exist and otherwise an update operation by default. When the newly inserted ${key}_{new}$, ${key}_n < {key}_{new} < {key}_{n+1}$, the data of ${key}_n$ or ${key}_{n+1}$ needs to be moved in order to write the data of ${key}_{new}$ to keep the data stored orderly.
\item \textbf{Data deletion.} If the corresponding key exists, it clears the corresponding indexing information and corresponding data.
\item \textbf{Data query.} In real business situations, the majority of data queries involve querying the data with a given key, or querying with a range of keys. For a query with a given key range, since the data is stored in the order of the key, the query function can be implemented very efficiently.
\end{itemize}

\subsection{Storage Model}
\begin{figure}[b]
	\centering
	\includegraphics[width=\linewidth]{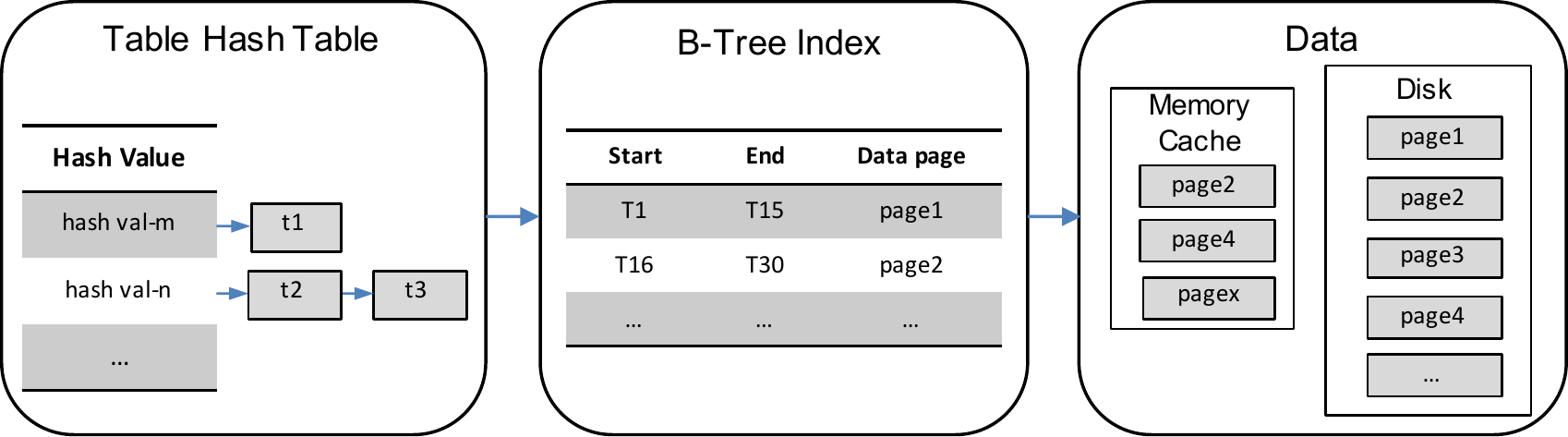}
	\caption{Storage module overview.}
	\label{fig:fig2}
\end{figure}
Brief overview on storage module in shown in Figure \ref{fig:fig2}. At the storage level, one database corresponds to a single file on the disk, and all the information contained in the database (such as table name, schema, indexing, user's KV data, etc) is stored in the disk file according to the page granularity, so as data access operations. 

The table information contained in the database is organized into strings by using a two-way linked list. The table information contains the file offset pointing to the root node of the table's B+ tree index, the leaf node of the table index contains the file offset pointing to user data.

When the storage engine reads or writes data from / to a specific table, it traverses the table list to get its information, and caches it into memory via a Hash Table (Table Hash Table). Finally, the file offset of the data can be quickly located by searching the table's B+ tree index.

To improve the efficiency of reading and updating on hot data, the storage engine introduces a memory cache, and the data modification corresponding to the Key occurs in the memory cache. If necessary, the data is written back to the corresponding file on disk. Compared with Log-based data writing and updating strategy, the Maxwell storage engine does not require subsequent processing after data is written, ensuring stable service continuity. By optimizing the memory cache policy, storage engine can obtain optimal access efficiency for specific scenarios. By controlling the cache size, storage engine can achieve a similar function to Redis distributed cache, with data persistence capabilities.

\subsection{Engine implementation}
Maxwell storage engine is designed and implemented from scratch fully utilizing the characteristics of disk, CPU, and operating system. The main technical points are summarized as follows:

\paragraph{Firstly,} Maxwell takes advantage of the characteristics of the SSD and abandons the LSM-Tree structure, all data is modified in-situ, and without compaction, which can keep the storage engine steadiness, reducing performance jitter, avoiding performance inflection. Mainstream storage engines, such as RocksDB \cite{rocksdb}, are essentially designed to optimize for HDD disks. The seek time of HDD is about 10ms, and disk IOPS is barely over one hundred, resulting in extremely low random access performance to disks. Therefore, the storage engine is designed to avoid random writes. The common practice is to use LSM-Tree \cite{gfs,cassandra} to convert discrete random write requests into batch sequential writes in MemCache to improve write performance, and then the read performance is optimized by Compaction. However, this brings in problems of read-write amplification and space enlargement. In addition, it will cause time-consuming fluctuations in the processing of read/write requests. For example, once Compaction or MemCache flushing occurs, it will greatly reduce the system's read/write throughput, and hurt the file system cache, thus causing performance jitters in storage systems and business links. This is the main reason for performance jitter and inflection. Optimizations \cite{raju2017pebblesdb, kim2019isolation, li2019elasticbf, zhang2020unikv, yue2016building, lee2021partial, mei2018sifrdb, balmau2019silk, conway2020splinterdb, yao2017light} were made to remedy the consequences of compaction. Maxwell chooses another way. Since SSD does not have issues such as seeking time, a single disk can achieve ~500'000 IOPS, its performance is far higher than HDD disks, which meets almost all daily business requirements. Therefore, Maxwell adopts the method of storing data in a page-by-page and in-situ modification without subsequent compaction.

\paragraph{Secondly,} Maxwell adopts a single-threaded work mode to reduce thread switching overhead, processes requests concurrently through coroutines, eliminates lock competition problems caused by data sharing between requests, and avoids performance degradation, jitter, and inflection in high-load scenarios. Unlike most storage engines, Maxwell does not use a thread pool to concurrently process read and write requests. All read/write requests are executed by a single-threaded time-sharing pipeline. The reason behind this is that the bus of the CPU is a thin copper film. The density of transistors per unit area increases according to Moore's Law, the delay caused by wire resistance and parasitic capacitance (RC delay) is rarely optimized. As a result, the computing unit needs to consume about 100ns and 1'000pJ power consumption to transform 8 bytes from main memory to CPU, while the computing unit only needs about 0.3ns and 10pJ power consumption to run. Therefore, reducing state changes and data migration are the keys to achieving high performance.
\newline Multithreading will increase OS context switching times, OS page cache misses, CPU jump instruction prediction failures, CPU cache misses, CPU execution channel conflicts, and etc. and will cause a large number of state switching and data migration, resulting in performance degradation. Coroutine is adopted to avoid those performance issues, although methods \cite{ma2017,ma2019} were proposed to optimize the efficiency of multi-threads. The cost of switching between coroutines is extremely low, avoiding the overhead of complex kernel mode and user mode switching, and the time overhead is reduced from microseconds to nanoseconds.
\newline Thread locks will severely reduce throughput and cannot achieve linear expansion. For example, if a task can be parallelized and has 64 CPU cores available, assuming that it contains only $ 1\%$ serialization code (such as Spinlock operation even if the performance reaches the theoretical value), then according to Amdahl's law, the throughput improvement: $1/ (0.99/64 + 0.01) = 39.26$ for 64 CPU cores. In other words, even using all 64 CPU cores, we can only achieve a throughput of fewer than 40 cores. This is also the main reason why some storage systems have performance inflection and system throughput doesn't scale linearly according to the number of resources used. Especially when task atomicity needs to be guaranteed, such as the aforementioned hot-spot accounting, the requests applied to the same account are naturally required to be processed one by one. Multi-threading does not help increase the processing throughput of hot-spot accounts, whereas the single-threaded working mode has significant advantages.

\paragraph{Thirdly,} Maxwell relies on a built-in autonomous cache management module to adjust the data cache replacement strategy according to the scene, adapt to different business needs, and to maximize the benefit of the available memory resources. This module also ensures that the data in memory is persisted on the disk.

\paragraph{Fourthly,} Maxwell uses a built-in network management module to realize ultra-low latency network packet sending, receiving and handling.

\paragraph{Fifthly,} Maxwell process all IO operations asynchronously. In this way, disk and network access can reach the theoretical bandwidth, and the request processing process does not wait or sleep to fully utilize the CPU computing power. Asynchronous IO can realize batch processing of IO operations with different requests, and further improve the processing efficiency of IO.

\paragraph{Sixthly,} Maxwell offers an optional user-mode file system based on Storage Performance Development Kit \cite{spdk} (SPDK) to further improve disk IO access efficiency, to reduce operating system impact and jitter, and to maximize performance in physical machine deployment scenarios.

\paragraph{Seventhly,} Maxwell is programmed by C language and assembly, with no third-party libraries used, no garbage collection, no implicit resource allocation and other operations. This kind of development policy is to minimize uncontrollable factors at the language level, and to maximize the use efficiency of software and hardware. In addition, it needs to be stressed that the Maxwell storage engine code is only hundreds of KB after being compiled, which can be used on platforms such as IoT and mobile devices.

\section{Compute Engine}
Maxwell's compute engine uses the single-process and single-thread mode in each node, so there is no competition overhead for synchronous locks between user requests and no thread switching overhead. Multiple compute engines form a master-slave distributed computing structure. The number of computing engines can be flexibly expanded. Each compute engine can access data from all storage engines. Compute engines and storage engines can be deployed on the same server or in a distributed manner. Compute engine can support millions of transaction processing capabilities and has linear scaling.

The compute engine provides the runtime environment for executing user requests. Once a user request is received by the compute engine, it creates a coroutine context for the request and starts to execute. Interfaces are provided for requests to access data on remote storage engines and local compute engines. The request can also create multiple sub-requests for local or remote calling to other computing engines for computing as needed, which implement distributed processing.

Request's processing logic totally depends on the business logic-flow. The flow of requests between compute engines and the mapping between user data to different storage engines can be customized to fulfill different business requirements. The compute engine provides user-defined functions to help users sink the upper-layer business logic, reduce data interaction between the client and the server, and improve execution efficiency. It also supports business development with Python.
\begin{figure}[h]
	\centering
	\includegraphics[width=\linewidth]{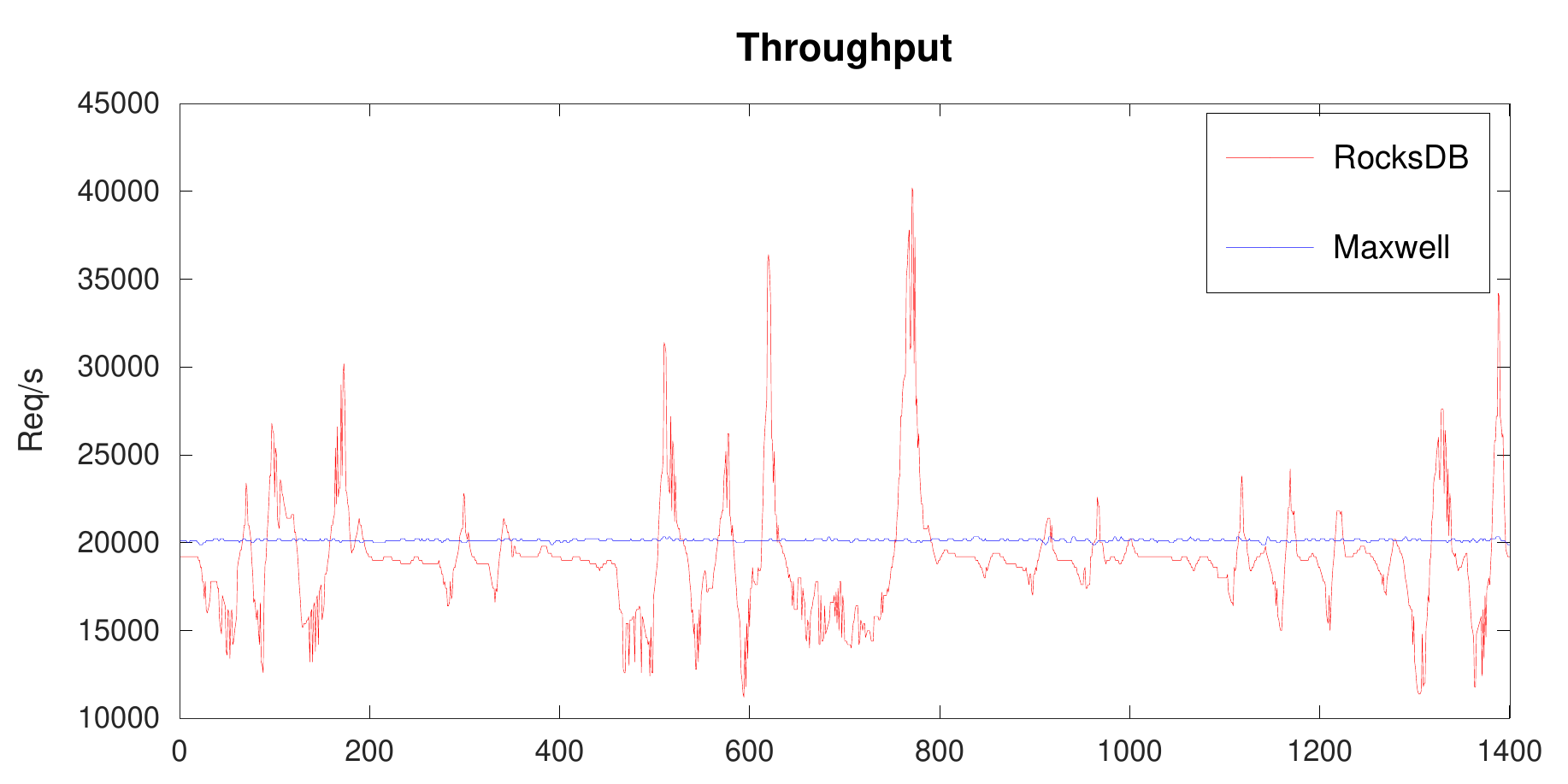}
	\caption{Average throughput of storage engines.}
	\label{fig:fig3}
\end{figure}
\begin{figure}[h]
	\centering
	\includegraphics[width=\linewidth]{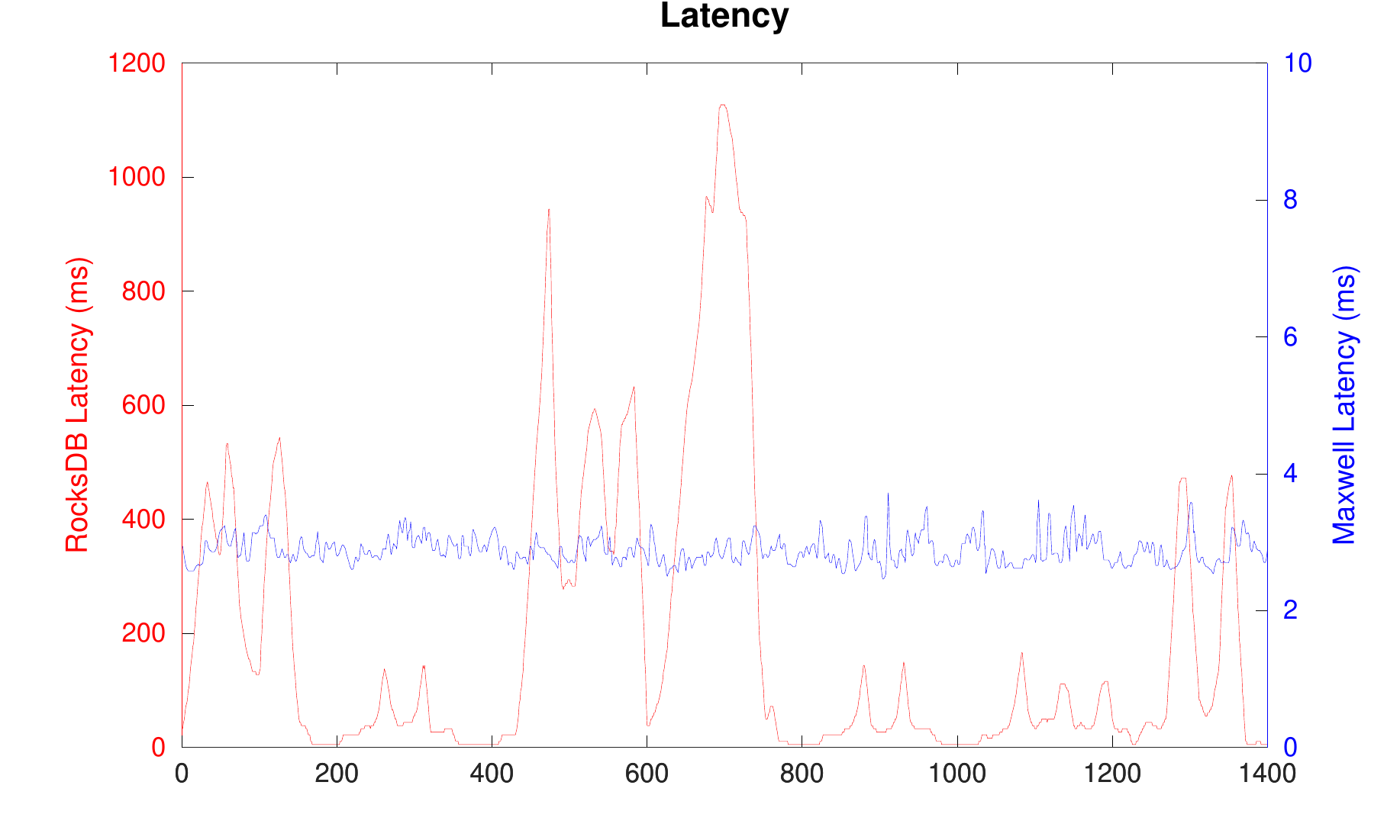}
	\caption{Average latency of storage engines.}
	\label{fig:fig4}
\end{figure}

\section{ARaft}
ARaft is a fully self-developed data synchronization module based on the Raft consistency protocol \cite{raft}. It also supports system upgrades and O \& M. ARaft uses a built-in Maxwell's storage engine to store Raft Log, which improves the efficiency of the system and eliminates the jitter caused by log storage. Based on single-thread mode. It merges client requests dynamically and adjusts the RLog distribution process adaptively. This helps to achieve high performance under both low traffic and heavy pressure scenarios, and to buffer the impact of network jitter between replicas.

ARaft provides strong and weak consistent query modes. For a strong consistent query, only the Leader responds to the query requests. The read and write requests are processed in sequence through Log distribution and execution. For weak consistency queries, read requests are directly sent to compute engines for processing without passing through Log distribution. Therefore, the order of read and write requests isn't guaranteed. In turn, all ARaft nodes can provide read services to increase system query throughput, which is widely used in many scenarios where data inconsistency can be tolerated in a short time, but with strict requirements on query throughput and overall path latency.

ARaft implements dynamic load balance based on the latency between clients and the ARaft service node to reduce the query time. It can self-heal when the network is broken or jitters. Clients periodically send detection messages to each ARaft node, calculate the time consumed when receiving the response, and select the node with the shortest time as the processing node for weak consistency queries.

The ARaft, compute engine, and storage engine are all pairwisely connected through sockets to form a logical replica. If one of the instances exits or the network is disconnected, the replica becomes unavailable immediately, ensuring that all replicas in the cluster are consistent and valid.

\begin{figure}[b!]
	\centering
	\includegraphics[width=\linewidth]{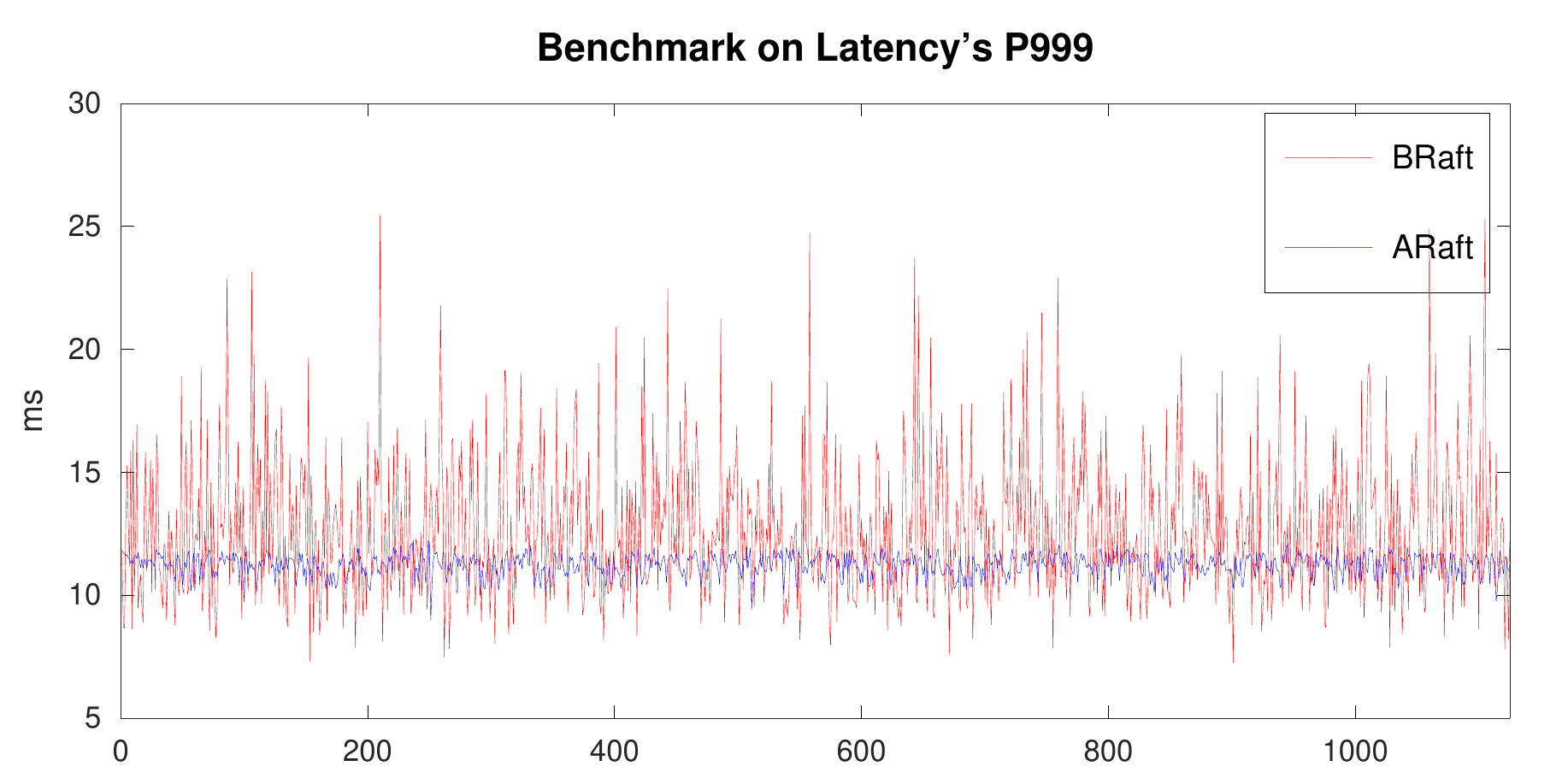}
	\caption{P999 latency distribution comparison between ARaft and BRaft.}
	\label{fig:fig5}
\end{figure}
\begin{figure}[b!]
	\centering
	\includegraphics[width=\linewidth]{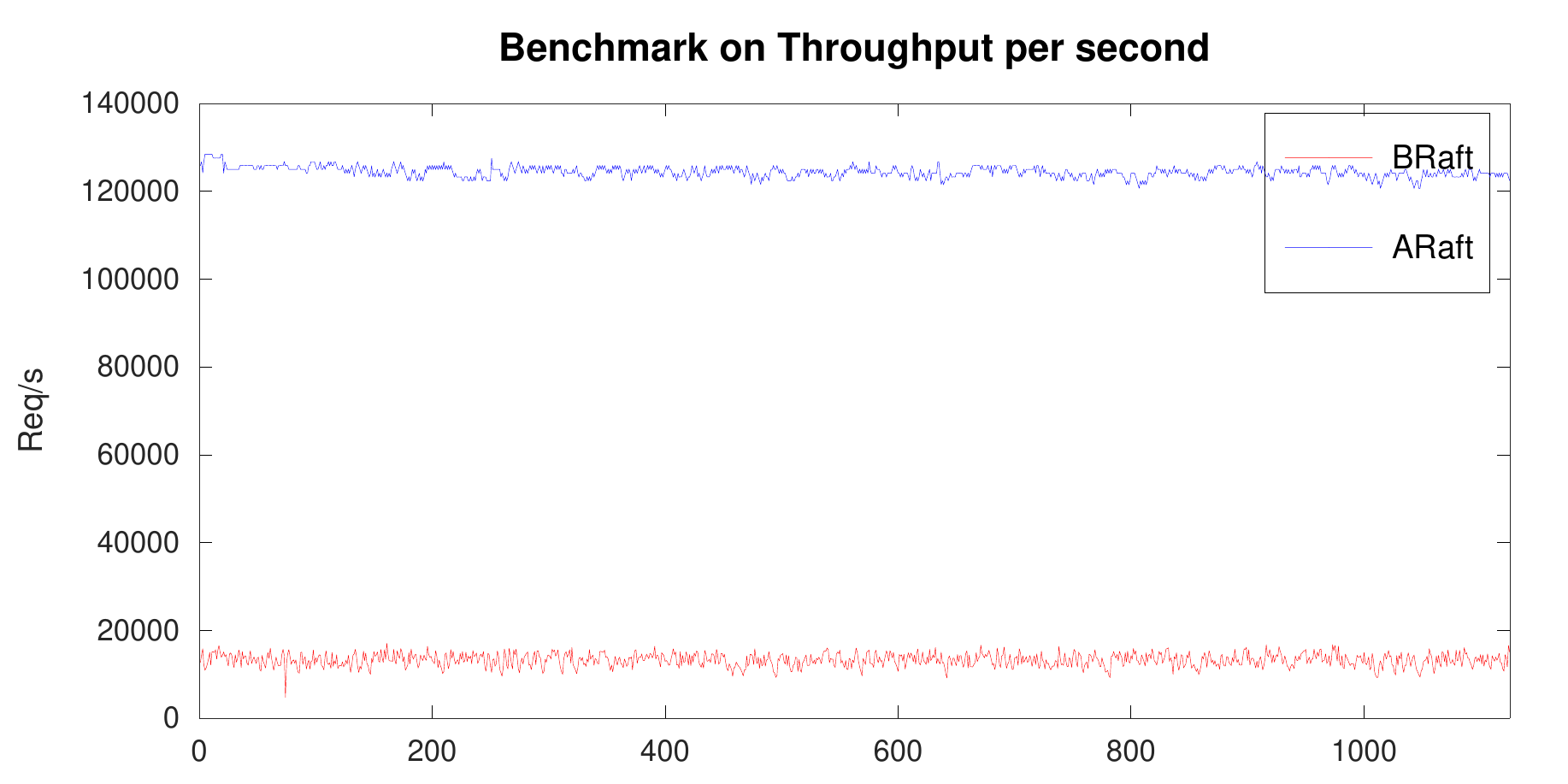}
	\caption{Throughput comparison between ARaft and BRaft.}
	\label{fig:fig6}
\end{figure}
\section{Evaluation}
\subsection{Compute-Storage Engine}
To evaluate the impact of different storage engines on system performance, evaluation follows the same test environment.
\begin{itemize}
\item \textbf{Test Method:} 100 clients send requests with a total pressure of 20'000 TPS on a cluster with three replicas.
\item \textbf{Test Machine:} Dockers with 4 CPU cores (Intel Xeon Platinum 8163 2.50 GHz CPU), 32 G RAM, and NVME SSD.
\end{itemize}

Figures \ref{fig:fig3} and \ref{fig:fig4} show the average system throughput and latency (i.e. request time consumed) by different storage engines, respectively. The accounting throughput of RocksDB fluctuates between 10'000 TPS and 40'000 TPS, and the average accounting time fluctuates dramatically from several ms to 1'200ms, which is intolerable for the hot-spot accounting system. In fact, it will lead to a series of consequences such as accounting failure and transaction failure.

In contrast, the accounting throughput Maxwell storage engine is basically stable at 20'000 TPS, and the average accounting cost is also stable, about 3ms.

\subsection{ARaft}
Compared with Braft, another widely used open-source Raft implementation, using the same test conditions, ARaft has the same average latency with BRaft. But as shown in Figure \ref{fig:fig5}, ARaft's P999 latency distribution is much more stable than BRaft. And the throughput of ARaft is $9.1$ times of BRaft, brief throughput comparison result is shown in Figure \ref{fig:fig6}.

\begin{figure}[h!]
	\centering
	\includegraphics[width=\linewidth]{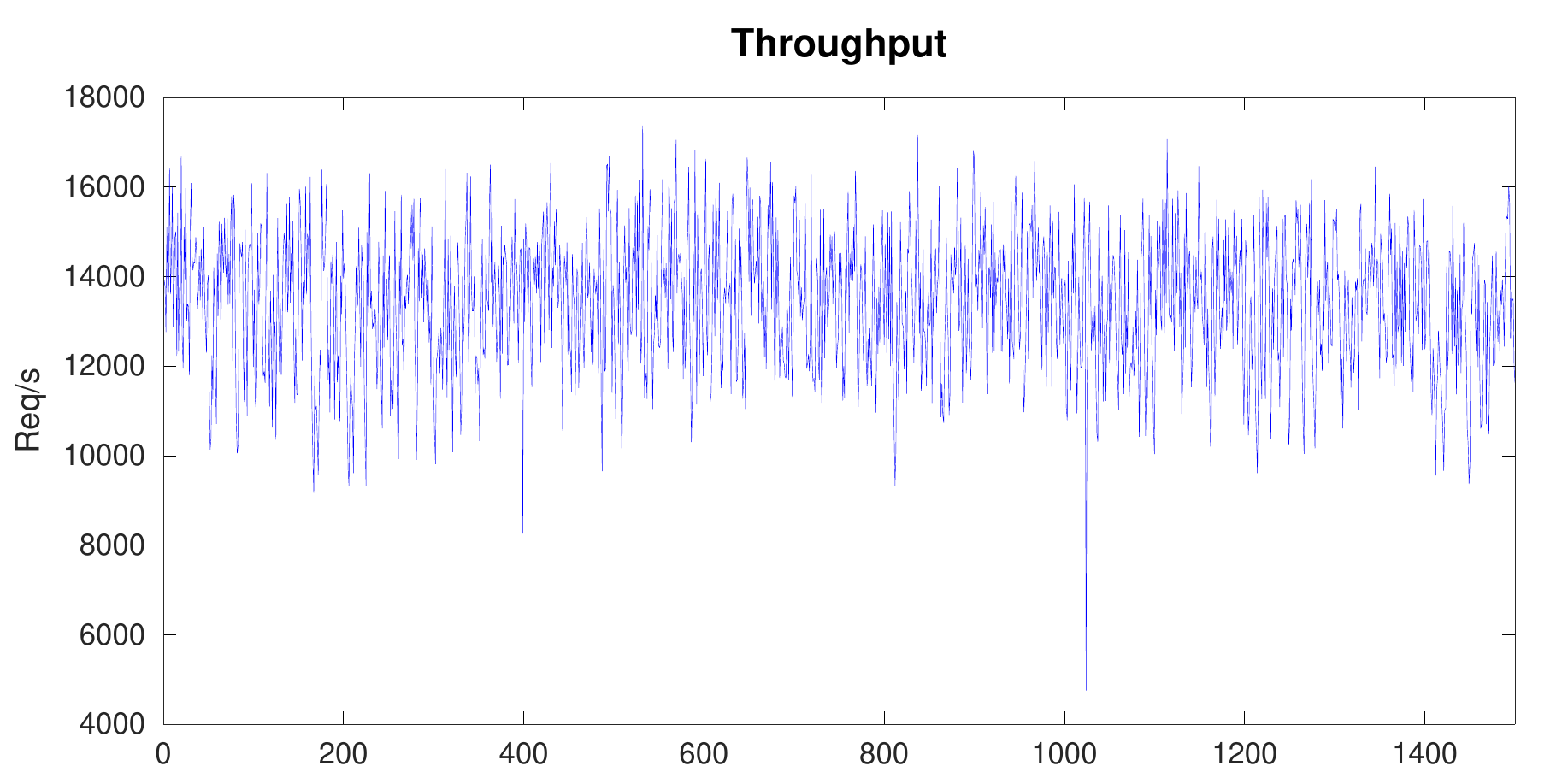}
	\caption{BRaft throughput drops when compacting.}
	\label{fig:fig7}
\end{figure}

\begin{figure}
	\centering
	\includegraphics[width=\linewidth]{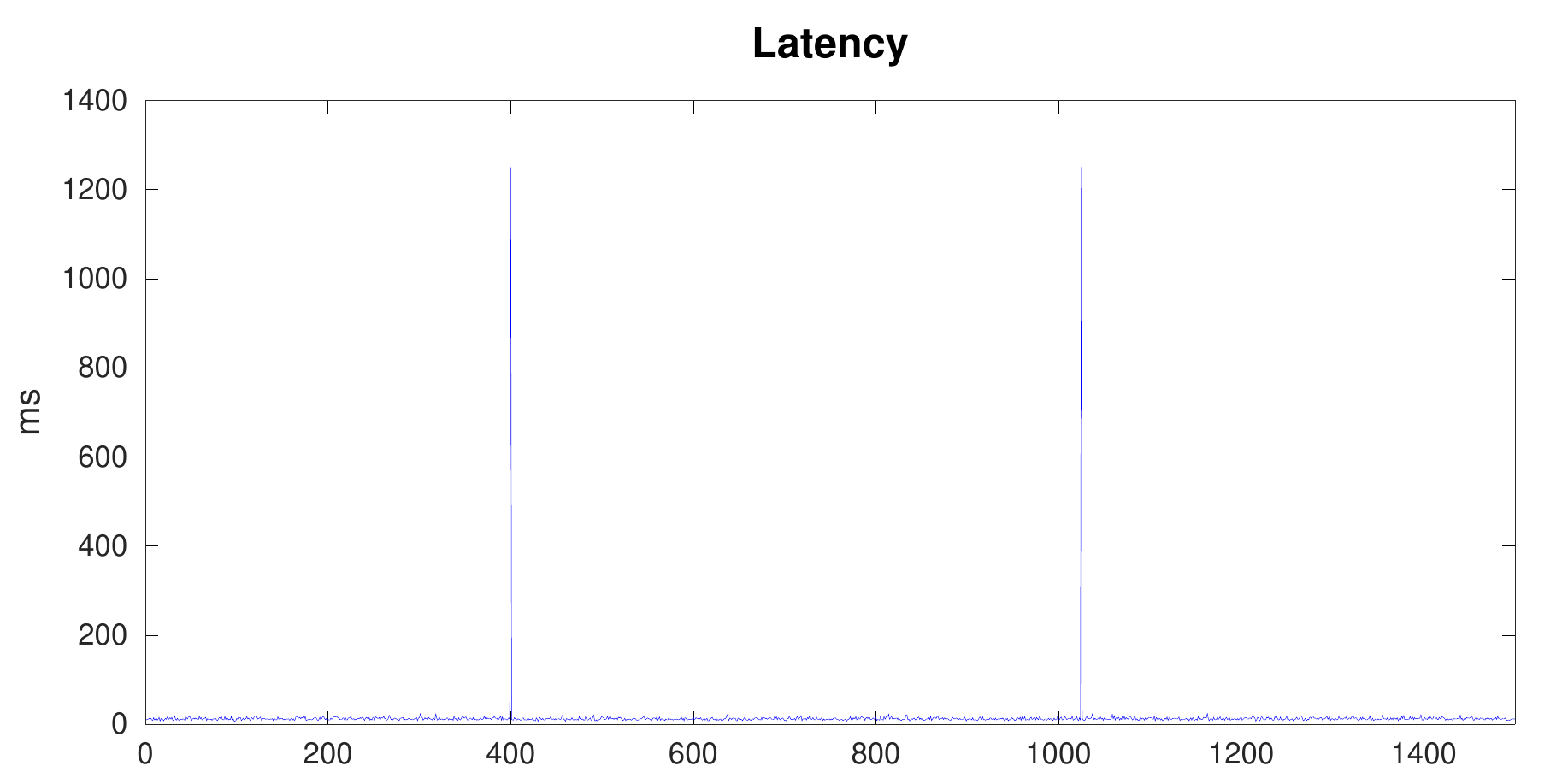}
	\caption{BRaft latency decays when compacting.}
	\label{fig:fig8}
\end{figure}

\begin{figure}
	\centering
	\includegraphics[width=\linewidth]{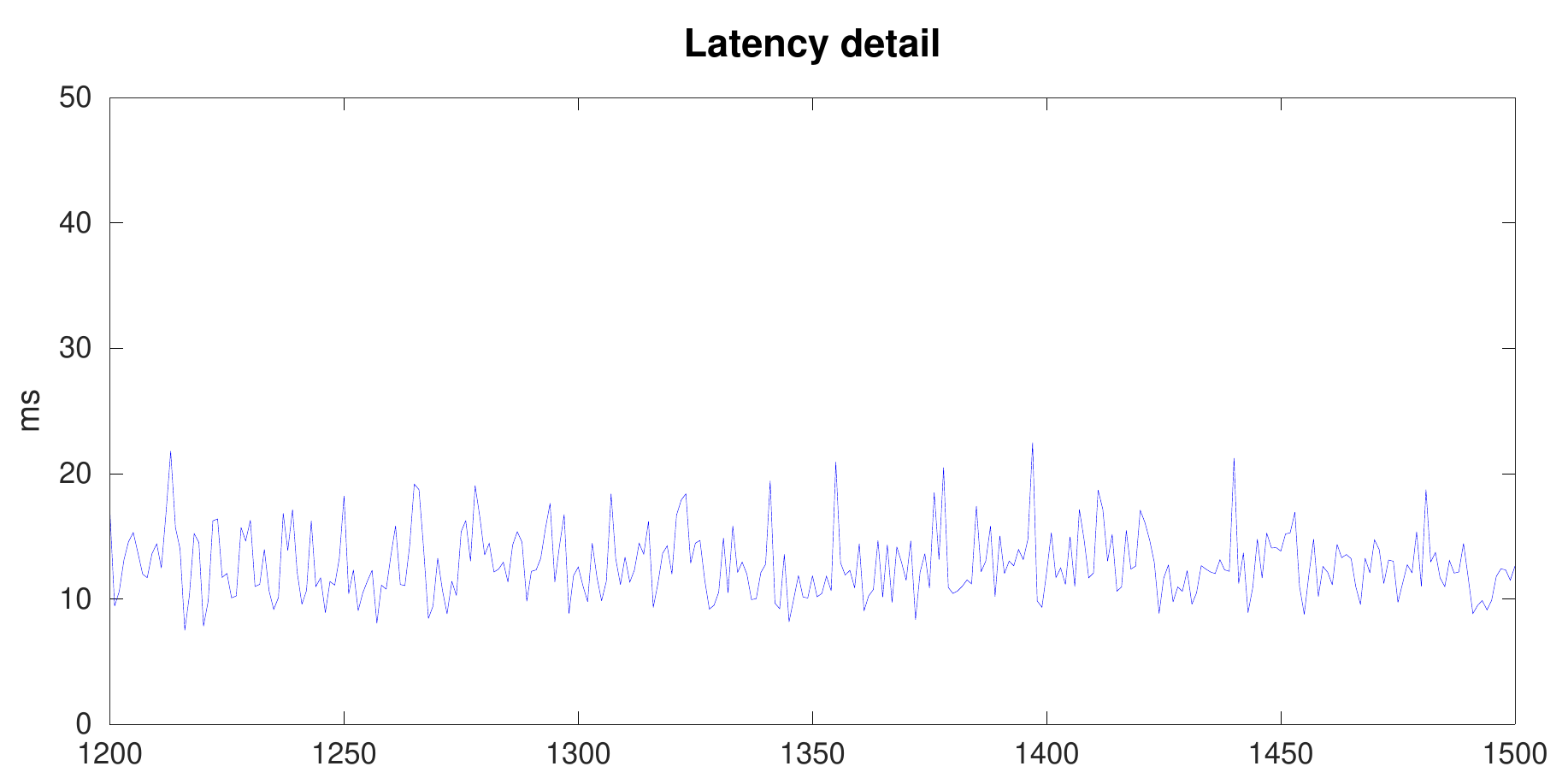}
	\caption{BRaft normal latency.}
	\label{fig:fig9}
\end{figure}

As the testing continuing on, compaction occurs in the open-source version. When compaction occurs, the request throughput decreases significantly (Figure \ref{fig:fig7}) and the latency increases sharply (Figure \ref{fig:fig8}), while its normal latency is shown in Figure \ref{fig:fig9}. But ARaft can maintain stable throughput and latency due to log stroage is performed by its embedded Maxwell storage engine.

\subsection{Deployment in Ant Group}
Maxwell compute-storage system has already been deployed in Alipay for production, with more than a thousand times improvement over the original system. Due to its good performance, Maxwell now carries hundreds of millions accounts, and covers several aspects of financial services in Ant Group. Maxwell helps on keeping good user experience, saving at least 100 million RMB per year. Through continuous iteration and accumulation, it has changed from a scenario-oriented dedicated system to a high-performance compute-storage framework that is able to serve more business scenarios. Maxwell supports to quickly build solutions to different business areas demanding extreme performance.


It is worth mentioning that the Maxwell compute-storage system has been adopted by many critical businesses to support the Double 11 shopping festival, and is spreading across the entire Ant Group.

\section{Conclusion}

Maxwell distributed compute-storage system helps to improve overall performance by hundreds of times, even up to three orders of magnitude in many scenarios. They are achieved by making full utilization of capabilities of existing disks, networks, operating systems, and CPUs and etc, without depending on any specific high-end hardware accelerators. We can say that Maxwell is the result of sophisticated integration of software and hardware knowledge. Maxwell is programmed by C and assembly languages, and can be easily customized and tailored. Maxwell can be deeply integrated with businesses and optimized to provide high reliability, high throughput with no inflection point and no Jitter, and low latency compute and storage solution, along with reduction of business costs from the perspectives of hardware resource and energy consumption.

We are looking forward to Maxwell's continued growth, heading to the efficient and collaborative operation of computer components, helping businesses in different scenarios reduce computing and storage costs, and delivering the ultimate performance on common hardware.

\bibliographystyle{ACM-Reference-Format}
\bibliography{Maxwell}

\end{document}